# Quasi-one-dimensional Ising models with defects of the "random local field" type: Imry-Ma phase in spaces with dimension higher than the lower critical one

A.A. Berzin[a], A.I. Morosov[b]*, and A.S. Sigov[a]

[a]MIREA - Russian Technological University, 78 Vernadskiy Ave., 119454 Moscow, Russian Federation

[b]Moscow Institute of Physics and Technology (National Research University), 9 Institutskiy per., 141700 Dolgoprudny, Moscow Region, Russian Federation

**\*** e-mail: mor-alexandr@yandex.ru

The phase diagram in coordinates "temperature - concentration of defects" of quasi-one-dimensional Ising models with defects of the "random local field" type is investigated. The confrontation of the tendency to the emergence of the long-range order due to a weak interaction between one-dimensional spin chains and the tendency to the formation of the Imry-Ma phase in which the order parameter follows the fluctuations of the random field created by defects is studied. The possibility of the appearance of the Imry-Ma phase in a situation where the space dimension exceeds the lower critical dimension is shown. The question of the existence of the long-range order in the Ising model with random fields in space with the critical dimension $d_l = 2$ is considered.

## 1. Introduction

A long-range order in the chain of Ising spins does not exist at temperature different from the absolute zero. With decreasing temperature in a defect-free system, the one-dimensional spin correlation radius increases exponentially. When it reaches its critical size, the weak exchange interaction of spins belonging to neighboring spin chains in a quasi-one-dimensional system becomes significant. The crossover occurs from the one-dimensional behavior to the $d$-dimensional ($d \geq 2$) one, and the long-range order arises in the system.

In the quasi-one-dimensional system of Ising spins with defects of the "random local field" type, another scenario of the system behavior may arise. As temperature $T$ decreases below a certain value of $T^*$, a transition occurs from dynamic fluctuations of the order parameter to its static fluctuations, which follow the random defect field fluctuations. As a result, the disordered Imry-Ma state arises, which persists up to the absolute zero.

The first or second scenario is realized, it depends on the relationship between the characteristic scale of static fluctuations, which depends on the concentration of defects, and the critical correlation radius, which is determined by the inter-chain interaction. This work is devoted to finding the areas of physical parameters that correspond to the indicated scenarios and constructing the phase diagram "temperature - concentration of defects" of the system under consideration.

## 2. Energy of classical Ising spins

The energy of the exchange interaction of classical Ising spins forming the $d$-dimensional square lattice in the approximation of the nearest neighbors interaction is equal to

$$W_{ex} = -\frac{1}{2} J \sum_{i,\delta} \sigma_i \sigma_{i+\delta}, \tag{1}$$

where $J > 0$ is the exchange integral, the spin projection on the easy axis takes the values ± 1, the summation over $i$ is carried out over the entire lattice of spins, and $\delta$ numbers the nearest neighbors of the given spin.

The energy of interaction of spins with random local fields of defects is

$$W_{def} = -\sum_l \sigma_l h_l, \qquad (2)$$

here the summation is performed over defects randomly located at the lattice sites, $h_l$ is the local field of the $l$-th defect which randomly takes the values $\pm h_0$.

### 3. The long-range order in the two-dimensional Ising model with defects of the "random local field" type

It is well known that the lower critical dimension of the random-field Ising model with a short-range exchange interaction is two [1-3]. At the same time, there are opposite points of view on the question of the existence of a long-range order for the space dimension $d_l = 2$. A negative answer was given to this question in Refs. [4, 5], while it was shown [6, 7] that in the region of weak random fields at zero temperature the long-range order takes place. It should be noted that the authors, as a rule, consider a square lattice of Ising spins in which a random field exists at each site of the lattice and its value is described by the Gaussian distribution.

Consider a more realistic model in which random fields are created by defects. We give for this case the arguments proposed by Imry and Ma [1]. The average (per a unit cell) field of defects in a region with a linear size $L$ (in units of the lattice constant), due to statistical fluctuations in the number of defects with the opposite direction of the random field, is $h_0 c^{1/2}/L^{d/2}$, where $c \ll 1$ is the dimensionless concentration of defects (their number per a unit cell) [8]. The gain in energy per a unit cell due to the appearance of the Imry-Ma phase in which the order parameter in each region of the size $L$ is directed along the average field is

$$w_{I-M} \approx -\frac{h_0 c^{1/2}}{L^{d/2}}. \qquad (3)$$

The loss in energy per a unit cell due to the appearance of sharp domain walls at the boundaries of the regions is

$$w_{ex} \approx \frac{2dJ}{L}. \tag{4}$$

Thus, for $d < 2$, the Imry-Ma phase becomes energetically favorable at a sufficiently large value of $L$. The optimum value $L^*$ corresponding to the minimum of the total energy $w_{I-M} + w_{ex}$, and the minimum value of energy $w^*$ itself for $d = 1$ are, respectively

$$L^* \approx \frac{16J^2}{ch_0^2}, \tag{5}$$

$$w^* \approx -\frac{ch_0^2}{8J}. \tag{6}$$

For $d = 2$, the dependence of $w_{I-M}$ and $w_{ex}$ on $L$ is the same. But for $J^2 \gg ch_0^2$, the long-range state turns out to be the ground one. This simple energy consideration is in agreement with the conclusions of Refs. [6, 7]. Consequently, the Imry-Ma phase is absent in the two-dimensional system, and therefore, in all quasi-two-dimensional systems in spaces with $d > 2$.

Let us consider the possibility of the existence of the Imry-Ma phase in the quasi-one-dimensional system with weak interaction between spin chains.

### 4. Quasi-one-dimensional Ising model with defects of the "random local field" type

Let the index $m$ number parallel spin chains forming a ($d$-1)-dimensional square lattice ($d \geq 2$) in a perpendicular section, and the index $i$ number the spins along this chain. Then the energy of interaction of spins has the form

$$W_{ex} = -J_\parallel \sum_{i,m} \sigma_{i,m}\sigma_{i+1,m} - \frac{1}{2}J_\perp \sum_{i,m,\delta} \sigma_{i,m}\sigma_{i,m+\delta}, \tag{7}$$

summation over $i$ and $m$ is carried out over the entire spin lattice, and over $\delta$ it is carried out over the nearest neighboring chains to the given one. The exchange interaction of neighboring spins belonging to the same chain $J_\parallel > 0$ far exceeds that between neighboring spins belonging to different chains $J_\perp > 0$. The interaction energy of spins with random local fields of defects is given by formula (2), if we assume that the index $l$ defines a pair of indices $i_l$, $m_l$.

The crossover temperature from one-dimensional to $d$-dimensional behavior and the appearance of the long-range order in the mean-field approximation is found from the condition that the susceptibility of the system of spins becomes infinite [9] or from the condition of equality of the temperature and energy of interaction of the correlated portion of spins, given by the one-dimensional correlation radius $r_\parallel$, with the molecular field

$$T \approx z J_\perp r_\parallel, \tag{8}$$

where $z$ is the number of spin chains nearest to the given one. In our model $z = 2(d-1)$. According to Ref. [9]

$$r_\parallel = \frac{1}{2} \exp\left(\frac{2 J_\parallel}{T}\right). \tag{9}$$

Substituting the expression for $r_\parallel$ into formula (8) and solving by iteration the resulting self-consistency equation for $T$, we obtain the temperature of the ferromagnetic transition in the defect-free system [9]

$$T_c \approx \frac{2 J_\parallel}{\ln\left(\frac{4 J_\parallel}{z J_\perp}\right)}. \tag{10}$$

The correction to the energy of the ground state of non-interacting defect-free spin chains due to the interaction between the chains and the appearance of the $d$-dimensional long-range order per a unit cell is

$$w_d = -\frac{z J_\perp}{2}. \tag{11}$$

Defects of the "random local field" type at $2 J_\parallel > h_0$ do not change the energy of the ordered state due to the random sign of the defect field.

If the Imry-Ma phase is realized in the system, then the correlation between the chains is violated, and the correction to the energy of the ground state of non-interacting defect-free spin chains due to interaction with random fields is described by formula (6) with $J \equiv J_\parallel$.

Thus, for the appearance of the Imry-Ma phase, it is necessary that

$$|w^*| > |w_d|, \qquad (12)$$

from whence we get the condition for the concentration of defects

$$c > \frac{4zJ_\perp J_\parallel}{h_0^2}. \qquad (13)$$

For $c \sim 10^{-2}$ and $h_0^2/J_\parallel^2 \sim 10^{-1}$ this gives the condition $J_\perp/J_\parallel \lesssim 10^{-4}$. In the case of such a weak exchange interaction, the dipole–dipole interaction between the spins can become significant. For the appearance of the Imry-Ma phase, its value likewise should not exceed $10^{-4}J_\parallel$.

Consequently, the Imry-Ma phase can be observed only in the case of a very weak interaction between spin chains. The temperature of its occurrence, that is, the temperature $T^*$ of the transition from dynamic fluctuations of the order parameter to its static fluctuations induced by large-scale fluctuations of the random field of defects, is found from the condition $L^* = r_\parallel$ [10]:

$$T^* \approx \frac{2J_\parallel}{\ln\left(\frac{32J_\parallel^2}{ch_0^2}\right)}. \qquad (14)$$

It is readily seen that the condition (12) is equivalent to the condition $T^* > T_c$.

The phase diagram of the quasi-one-dimensional Ising model with defects of the "random local field" type in the variables "temperature - concentration of defects" is shown in Figure.

It is well known [3] that a non-magnetic substitution impurity or vacancy in a two-sublattice collinear antiferromagnet in an external magnetic field is a "random local field" type defect. The value of the local field is directly proportional to the induction of the magnetic field applied collinearly to the magnetizations of sublattices. The Imry-Ma state can occur in the collinear phase of a quasi-one-dimensional antiferromagnet if the condition (13) is met.

## 5. Conclusion

The main conclusions of the work can be stated as follows:

1.  In the case of a low concentration of defects of the "random local field" type, for the lower critical dimension $d_l = 2$, in the model with the same magnitude exchange interaction with all nearest neighbors, the ground state is the state with the long-range order.

2.  In the case of the quasi-one-dimensional Ising model with defects of the "random local field" type, the disordered Imry-Ma phase may occur despite the fact that the low-temperature behavior of the system is effectively $d$-dimensional with $d \geq 2$, that is, the space dimension can exceed the lower critical dimension obtained for the case of a uniform magnitude of the exchange interaction with all nearest neighbors.

3.  For the occurrence of the Imry-Ma phase, it is necessary that the exchange between the spins of adjacent spin chains be four (or more) orders of magnitude weaker than the exchange of spins in the chain.

**Figure caption**

Phase diagram of the quasi-one-dimensional Ising model with defects of the "random local field" type for $z=4$, $h_0^2/J_\parallel^2 = 10^{-1}$, $J_\perp/J_\parallel = 10^{-4}$: *P* denotes paramagnetic phase, *F* denotes ferromagnetic phase, and *I-M* is the Imry-Ma phase.

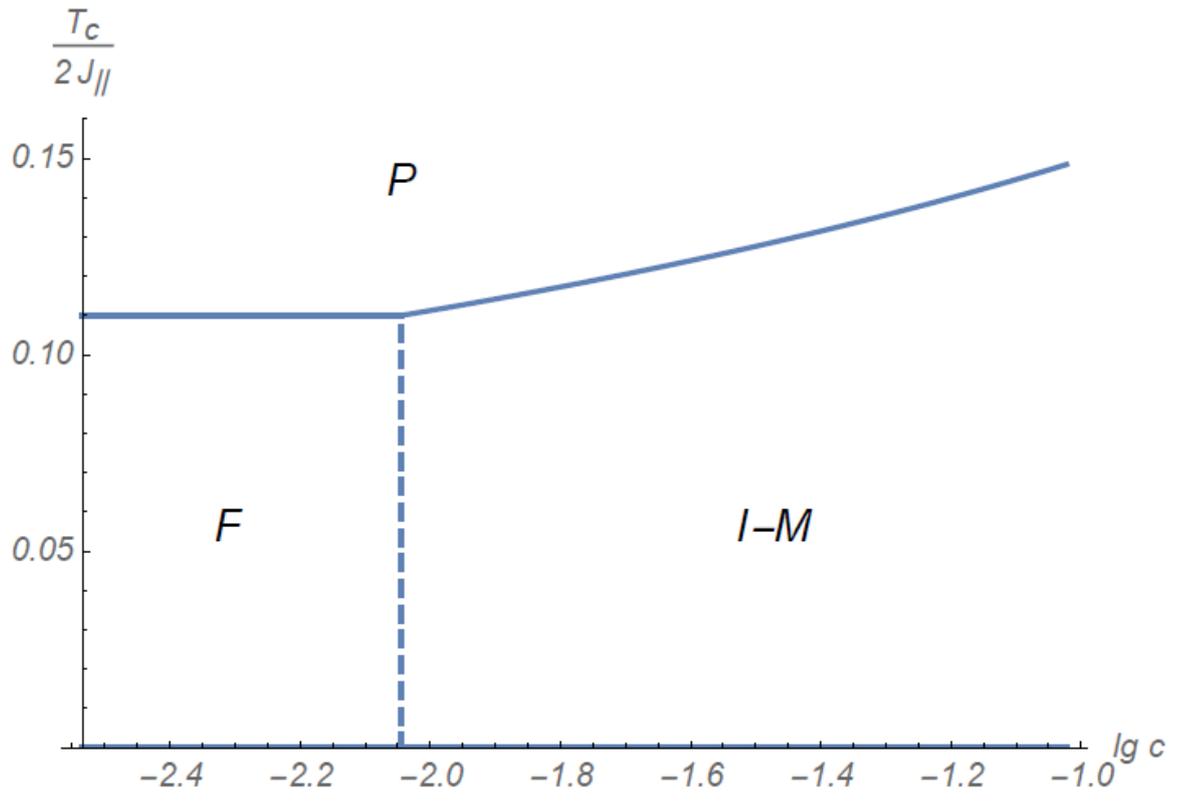